\documentclass[3p]{elsarticle}

\usepackage[utf8]{inputenc} % allow utf-8 input
\usepackage[T1]{fontenc}    % use 8-bit T1 fonts
\usepackage{xcolor}
\usepackage{hyperref}       % hyperlinks
\usepackage{url}            % simple URL typesetting
\usepackage{booktabs}       % professional-quality tables
\usepackage{amsfonts}       % blackboard math symbols
\usepackage{nicefrac}       % compact symbols for 1/2, etc.
\usepackage{microtype}      % microtypography
\usepackage{lipsum}
\usepackage{physics}
\usepackage{mathtools}
\usepackage{amsmath}
\usepackage{xcolor}
\usepackage{fancyhdr}       % header
\usepackage{graphicx}       % graphics
\graphicspath{{media/}}     % organize your images and other figures under media/ folder

\biboptions{sort&compress}

\makeatletter
\def\ps@pprintTitle{%
  \let\@oddhead\@empty
  \let\@evenhead\@empty
  \def\@oddfoot{\reset@font\hfil\thepage\hfil}
  \let\@evenfoot\@oddfoot
}
\makeatother
\begin{document}

\begin{frontmatter}

\title{Classical-to-quantum convolutional neural network transfer learning}

% \address{Radarweg 29, Amsterdam}
% \fntext[myfootnote]{Since 1880.}

\author[1]{Juhyeon Kim}
\address[1]{
SKKU Advanced Institute of Nanotechnology, Sungkyunkwan University, Suwon, Republic of Korea
}

\author[1,2,3]{Joonsuk Huh}
\ead{joonsukhuh@gmail.com}
\address[2]{
Department of Chemistry, Sungkyunkwan University, Suwon, Republic of Korea
}
\address[3]{
Institute of Quantum Biophysics, Sungkyunkwan University, Suwon, Republic of Korea
}

\author[4,5]{Daniel K. Park\corref{corr1}}
\ead{dkd.park@yonsei.ac.kr}
\address[4]{
Department of Applied Statistics, Yonsei University, Seoul, Republic of Korea
}
\address[5]{
Department of Statistics and Data Science, Yonsei University, Seoul, Republic of Korea
}

\cortext[corr1]{Corresponding author}

% \author[mymainaddress,mysecondaryaddress]{Elsevier Inc}

% \author[mysecondaryaddress]{Global Customer Service\corref{mycorrespondingauthor}}
% \cortext[mycorrespondingauthor]{Corresponding author}

% \address[mymainaddress]{1600 John F Kennedy Boulevard, Philadelphia}
% \address[mysecondaryaddress]{360 Park Avenue South, New York}

\begin{abstract}
Machine learning using quantum convolutional neural networks (QCNNs) has demonstrated success in both quantum and classical data classification. In previous studies, QCNNs attained a higher classification accuracy than their classical counterparts under the same training conditions in the few-parameter regime. However, the general performance of large-scale quantum models is difficult to examine because of the limited size of quantum circuits, which can be reliably implemented in the near future. We propose transfer learning as an effective strategy for utilizing small QCNNs in the noisy intermediate-scale quantum era to the full extent. 
In the classical-to-quantum transfer learning framework, a QCNN can solve complex classification problems without requiring a large-scale quantum circuit by utilizing a pre-trained classical convolutional neural network (CNN). We perform numerical simulations of QCNN models with various sets of quantum convolution and pooling operations for MNIST data classification under transfer learning, in which a classical CNN is trained with Fashion-MNIST data. The results show that transfer learning from classical to quantum CNN performs considerably better than purely classical transfer learning models under similar training conditions.
\end{abstract}


\end{frontmatter}

\section{Introduction}
\label{sec:intro}
Machine learning (ML) with a parameterized quantum circuit (PQC) is a promising approach for improving existing methods beyond classical capabilities~\cite{Romero_2017,PhysRevA.98.032309,Benedetti_2019,cong_quantum_2019,cerezo2020variational,mangini_quantum_2021,PRXQuantum.2.040337}. This is a classical-quantum hybrid algorithm in which the cost function and its corresponding gradient are computed using quantum circuits~\cite{PhysRevLett.118.150503,PhysRevA.99.032331} and the model parameters are updated classically. Such hybrid ML models are particularly advantageous when cost function minimization is difficult to perform classically~\cite{peruzzo_variational_2014,McClean_2016,cong_quantum_2019}. These models optimize the quantum gate parameters under the given experimental setup, and hence can be robust to systematic errors. Furthermore, they are less prone to decoherence because iterative computation can be exploited to reduce the quantum circuit depth. Thus, the hybrid algorithm has the potential to achieve quantum advantage in solving various problems in the noisy intermediate-scale quantum (NISQ)\footnote{NISQ refers to the domain of quantum computing where the number of quantum processors that can be manipulated reliably is limited due to noise, yet holds the potential to surpasses the classical capabilities to a certain extent. As NISQ technology becomes increasingly accessible, the discovery of its real-world applications has become crucially important.} era~\cite{Preskill2018quantumcomputingin,bharti2021noisy}.

A critical challenge in the utilization of PQC for solving real-world problems is the barren plateau phenomenon in the optimization landscape, which makes training the quantum model that samples from the Haar measure difficult as the number of qubits increases~\cite{McClean2018}. One way to avoid the barren plateau is to adopt a hierarchical structure~\cite{grant_hierarchical_2018,pesah2020absence}, in which the number of qubits decreases exponentially with quantum circuit depth, such as in quantum convolutional neural networks (QCNNs)~\cite{cong_quantum_2019}. The hierarchical structure is interesting from a theoretical perspective because of its close connection to tensor networks~\cite{grant_hierarchical_2018,HUANG202189}. Moreover, the shallow depth of a QCNN, which grows logarithmically with the number of input qubits, makes it well suited for NISQ computing. In addition, an information-theoretic analysis shows that the QCNN architecture can help reduce the generalization error~\cite{PRXQuantum.2.040321}, which is one of the central goals of machine learning. All these factors motivate the application of QCNN for machine learning. QCNNs have been shown to be useful for solving both quantum~\cite{cong_quantum_2019,maccormack_branching_2020} and classical~\cite{hur2021quantum} problems despite their restricted structure with a shallow-depth quantum circuit. In Ref.~\cite{hur2021quantum}, for binary classification with the MNIST~\cite{726791} and Fashion-MNIST~\cite{xiao2017fashionmnist} datasets, QCNN yielded higher classification accuracy than the classical convolutional neural network (CNN) when only 51 or fewer parameters were used to construct these models. The best-known classical CNN-based classifiers for the same datasets typically employ millions of parameters. However, the size of the quantum circuits that can be implemented with current quantum devices is too small to incorporate such a large number of parameters. Therefore, two important issues remain. The first is to verify whether a QCNN can continue to outperform its classical counterpart as the number of trainable model parameters increases. The second is to utilize small QCNNs that can be realized in the near future to the full extent, so that a quantum advantage can be achieved in solving practical problems. The latter is the main focus of this work.

An ML problem for which the quantum advantage in the few-parameter regime can be exploited is transfer learning (TL)~\cite{Bozinovski2020ReminderOT,5288526,10.1007/978-3-030-01424-7_27,Goodfellow-et-al-2016}. TL aims to utilize what has been learned in one setting to improve generalization in another setting that is independent of the former. TL can be applied to classical-quantum hybrid networks such that the parameters learned for a classical model are transferred to training a quantum model or vice versa~\cite{Mari2020transferlearning}. In the classical-to-quantum (C2Q) TL scheme, the number of qubits increases with the number of output nodes (or features) of the pre-trained classical neural network. This indicates that the transferred part of a classical neural network should have a small number of output nodes to find applications in the NISQ era. For example, using a pre-trained feedforward neural network with a large number of nodes throughout the layers would not be well suited for near-term hybrid TL. By contrast, building a TL model with a classical and quantum CNN is viable because the number of features in the CNN progressively decreases via subsampling (i.e., pooling), and the QCNN has already exhibited an advantage with a small number of input qubits.

Motivated by the aforementioned observations, we propose a TL framework for classical-to-quantum convolutional neural networks (C2Q-CNNs). Unlike previous works, C2Q-CNN transfers knowledge from a pre-trained (source) classical CNN to a quantum CNN, thereby preserving the benefits of quantum CNN. Our method avoids the need for classical data dimensionality reduction, commonly required in existing methods, as the classical CNN serves as the source for TL. Additionally, we introduce new ansatzes for both quantum pooling and convolutional operations, enriching the model selection in the QCNN.
To evaluate the performance of C2Q-CNN, we conduct numerical simulations on the MNIST data classification task using PennyLane~\cite{bergholm2020pennylane}. The classical CNN is pre-trained on the Fashion-MNIST dataset. The simulations assess the classification accuracy under different quantum convolution and pooling operations and compare C2Q-CNN with various classical-to-classical CNN (C2C-CNN) TL schemes. The results show that C2Q-CNN outperforms C2C-CNN with respect to classification accuracy under similar training conditions. Furthermore, the new quantum pooling operation developed in this work is more effective in demonstrating the quantum advantage.
% This article contributes to the existing knowledge in three ways. Firstly, we introduce the C2Q-CNNs transfer learning approach, which has the potential to achieve a quantum advantage in practical problems. Secondly, novel ansatzes for quantum pooling and convolutional operations are proposed to expand the selection of models available for QCNNs. Finally, a numerical simulation is conducted to evaluate the performance of C2Q-CNNs and is compared to that of C2C-CNNs, providing evidence of the potential quantum advantage.

The remainder of this paper is organized as follows. Section~\ref{sec:preliminaries} reviews QCNNs and TL. This section also introduces the generalization of the pooling operation of the QCNN. Section~\ref{sec:c2q} explains the general framework for C2Q-CNN TL. The simulation results are presented in section~\ref{sec:results}. MNIST data classification was performed with a CNN pre-trained for Fashion-MNIST data, and the performance of the C2Q-CNN models was compared with that of various C2C-CNN models. The conclusions and outlook are presented in Section~\ref{sec:conclusion}.

\section{Preliminaries}
\label{sec:preliminaries}
\subsection{Quantum convolutional neural network}
\label{sec:qcnn}
\begin{figure}[t]
    \centering
    \includegraphics[width=0.9\textwidth]{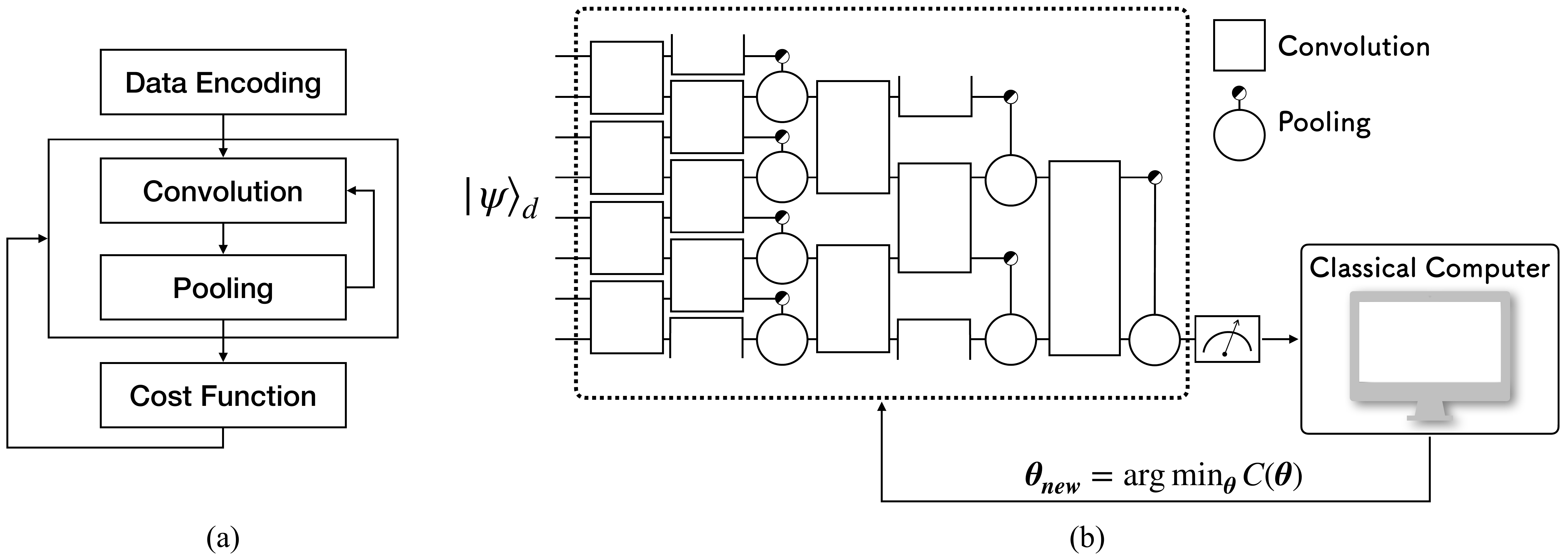}
    \caption{(a) Schematics of the QCNN algorithm with (b) an example for eight input qubits. Given a quantum state, $|\psi\rangle_{d}$, which encodes classical data, the quantum circuit comprises two parts: convolutional filters (rectangles) and pooling (circles). The convolutional filter and pooling use parameterized quantum gates. Three layers of convolution--pooling pairs are presented in this example. In each layer, the convolutional filter applies the identical two-qubit ansatz to the nearest neighbor qubits in a translationally invariant manner. The quantum convolutional operations in the QCNN circuits are designed to meet the closed boundary condition, as indicated by the open-ended gates in the figure, ensuring the top and bottom qubits in each layer are connected. Pooling operations within the layer are identical to each other, but differ from convolutional filters. The pooling operation is represented as a controlled unitary transformation, and the half-filled circle on the control qubit indicates that different unitary gates can be applied to each subspace of the control qubit. The measurement outcome of the quantum circuit is used to calculate the user-defined cost function. A classical computer is used to compute the new set of parameters based on the gradient, and the quantum circuit parameters are updated for the subsequent round. The optimization process is iterated until pre-selected conditions are met.}
    \label{fig:1}
\end{figure}

Quantum convolutional neural networks are parameterized quantum circuits with unique structures inspired by classical CNNs~\cite{cong_quantum_2019,hur2021quantum}. In general, QCNNs follow two basic principles of classical CNNs: translational invariance of convolutional operations and dimensionality reduction via pooling. However, QCNNs differ from classical CNNs in several aspects. First, the data are defined in quantum Hilbert space, which grows exponentially with the number of qubits. Consequently, the quantum convolutional operation is not an inner product, as in the classical case, but a unitary transformation of a state vector, which is a linear map that transforms a vector to a vector, whereas a classical convolution operation is a linear map that transforms a vector to a scalar. The pooling in QCNN traces out half of the qubits, similar to the pooling in the CNN that subsamples the feature space. Typically, the pooling layer includes parameterized two-qubit controlled-unitary gates, and the control qubits are traced out after the gate operations.
Without loss of generality, we refer to the structure of a parameterized unitary operator for either convolution or pooling as ansatz. The cost function of a model with given parameters is defined with an expectation value of some observable with respect to the final quantum state obtained after repeating quantum convolutional and pooling operations. The QCNN is trained by updating the model parameters to minimize the cost function until a pre-determined convergence condition is met.
The general concept of a QCNN is illustrated in Fig.~\ref{fig:1} (a). An example of a circuit with eight input qubits is shown in (b). The depth of the QCNN circuit after repeating the convolution and pooling until one qubit remains is $O(\log N)$, where $N$ is the number of input qubits. This shallow depth allows the QCNN to perform well on quantum hardware that will be developed in the near future.

\begin{figure}[t]
    \centering
    \includegraphics[width=1\textwidth]{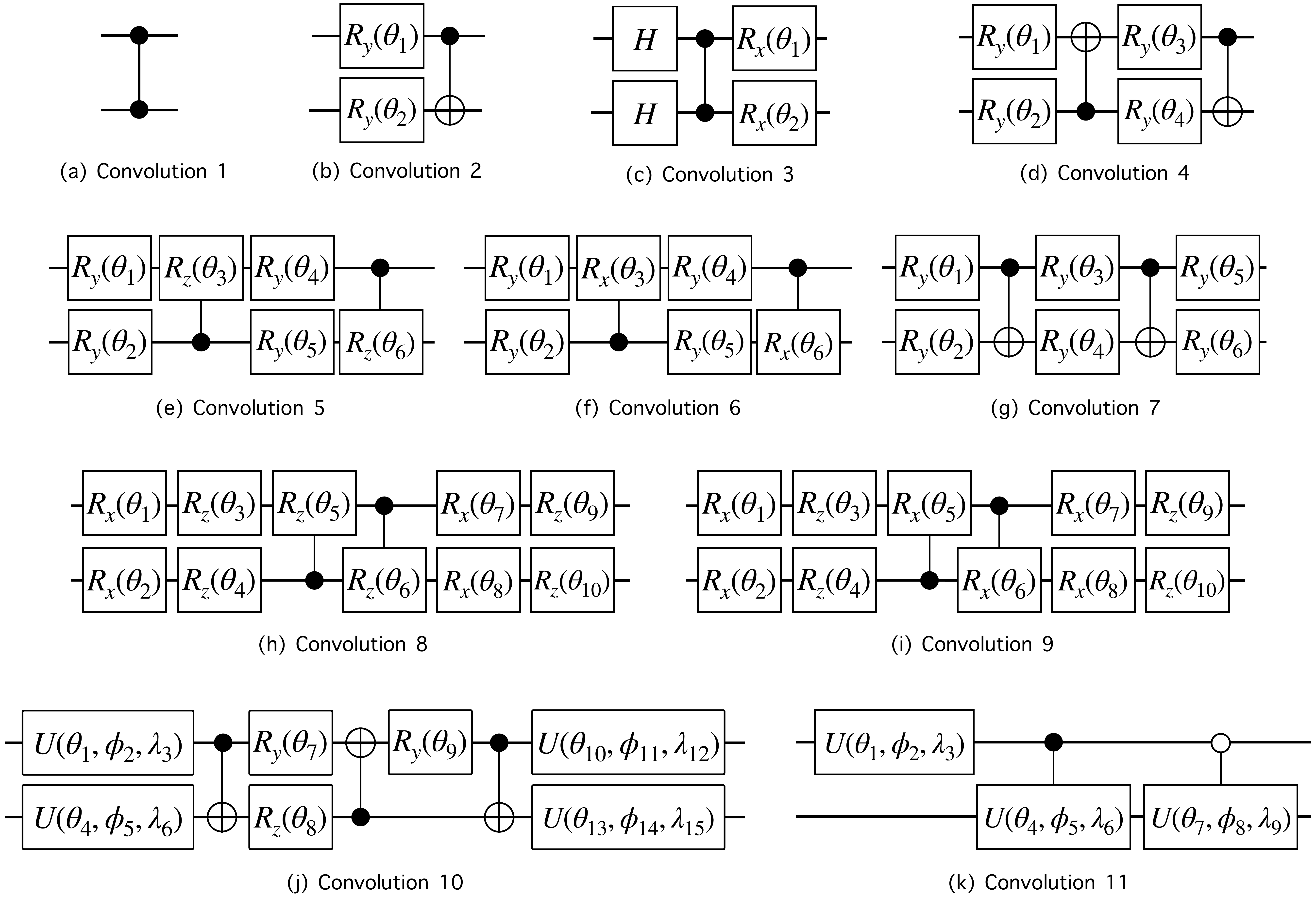}
    \caption{Parameterized quantum circuits used in the convolutional layer. The convolutional circuits from (b) to (j) are adapted from Ref.~\cite{hur2021quantum}, whereas (a) and (k) are the new convolutional circuits tested in this study.  $R_i(\theta)$ is the rotation around the $i$-axis of the Bloch sphere by an angle of $\theta$, and $H$ is the Hadamard gate. $U(\theta,\phi,\lambda)$ is an arbitrary single-qubit gate, which can be expressed as $U(\theta,\phi,\lambda) = R_z(\phi)R_x(-\pi/2)R_z(\theta)R_x(\pi/2)R_z(\lambda)$. $U(\theta,\phi,\lambda)$ can implement any unitary operation in $SU(2)$. As (j) can express an arbitrary two-qubit unitary gate, we test it without any parameterized gates for pooling in addition to ZX pooling and generalized pooling. For (k), we do not apply parameterized gates for pooling. In these cases, pooling simply traces out the top qubit after convolution.}
    \label{fig:convolution}
\end{figure}

The quantum convolution and pooling operations can be parameterized in many ways. 
The convolution ansatzes evaluated in this study are illustrated in Fig.~\ref{fig:convolution}. Among them, circuits (b) to (j) are the nine ansatzes previously tested in Ref.~\cite{hur2021quantum}. 
These ansatzes are motivated by past studies. For instance, circuit (b) is a parameterized quantum circuit that was used to train a tree tensor network (TTN)~\cite{grant_hierarchical_2018}.
The four-qubit parameterized quantum circuits analyzed by Sim et al.\cite{Sim_2019_expressibility} were modified to two-qubit circuits to serve as building blocks for the convolution layer, resulting in circuits (c), (d), (e), (f), (h), and (i). Circuits (h) and (i) are two-qubit versions of the circuits with the best expressibility, while circuit (c) is of the best entangling circuit. Circuits (d), (e), and (f) represent a good balance of expressibility and entangling capability.
Circuit (g) is used for the two-body variational quantum eigensolver entangler~\cite{parrish_2019} and can generate arbitrary $SO(4)$ gates~\cite{weidi_2012}, making it a suitable candidate for building convolution layer in QCNN. Circuit (j) is a parameterized arbitrary $SU(4)$ gate~\cite{vatan_2004, maccormack_branching_2020} capable of performing arbitrary two-qubit unitary operations.
Because the convolutional operations act on two qubits, parameterized $SU(4)$ operations provides the most general ansatz.
In this study, we introduce two new convolutional ansatzes, (a) and (k), to our benchmark. The former aims to study the classification capability of a QCNN when only pooling operations are trained. The latter is inspired by the generalized pooling operation described in the following paragraph, with an $SU(2)$ gate applied to a control qubit to split the subspaces in an arbitrary superposition.

\begin{figure}[t]
    \centering
    \includegraphics[width=0.65\textwidth]{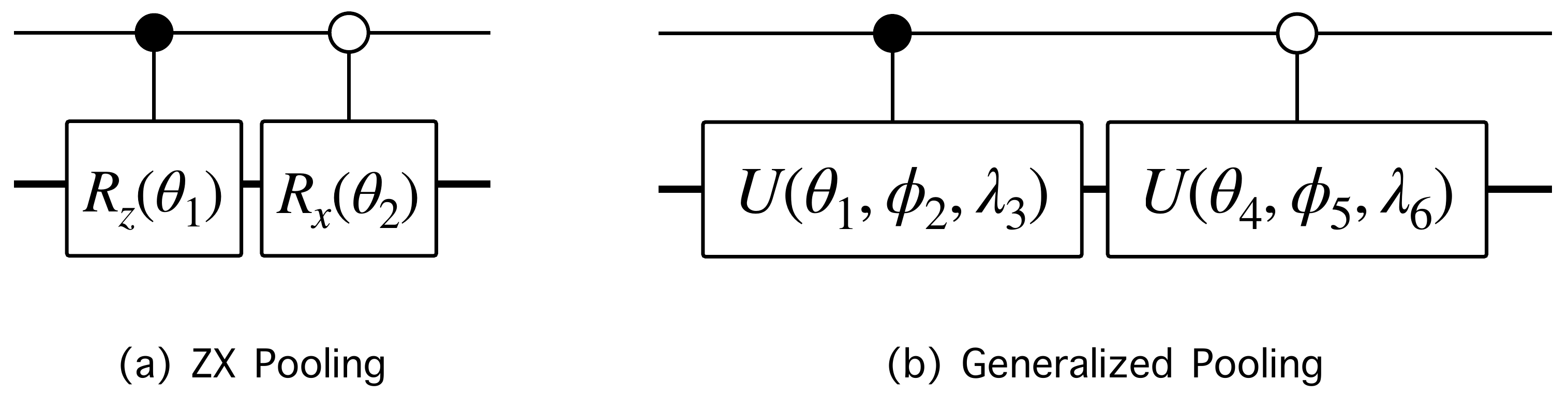}
    \caption{Parameterized quantum gates used in the pooling layer. The pooling circuit (a) is adapted from Ref.~\cite{hur2021quantum}, and (b) is the generalized pooling method introduced in this work. Generalized pooling applies two arbitrary single-qubit unitary gate rotations, $U(\theta_1,\phi_2,\lambda_3)$ and $U(\theta_4,\phi_5,\lambda_6)$, which are activated when the control qubit is 1 (filled circle) or 0 (open circle), respectively. The control (first) qubit is traced out after the gate operations to reduce the dimensions. The single-qubit unitary gate is defined as $U(\theta,\phi,\lambda) = R_z(\phi)R_x(-\pi/2)R_z(\theta)R_x(\pi/2)R_z(\lambda)$, and it can implement any unitary in $SU(2)$. The thinner horizontal line (top qubit) indicates the qubit that is being traced out after gate operations. }
    \label{fig:pooling}
\end{figure}

The pooling ansatzes used in previous studies were simple single-qubit-controlled rotations followed by tracing out the control qubit. For example, in Ref.~\cite{hur2021quantum}, a pooling operation in the following form was used:
\begin{equation}
\label{eq:ZXpooling}
    \Tr_A\left\lbrack\left(\ketbra{1}{1}_A\otimes R_z(\theta_1)_B + \ketbra{0}{0}_A\otimes R_x(\theta_2)_B\right) \rho_{AB} U^{\dagger}_{p} \right\rbrack,
\end{equation}
where $\Tr_A\left( \cdot \right)$ represents a partial trace over subsystem $A$, $R_i(\theta)$ is the rotation around the $i$ axis of the Bloch sphere by an angle of $\theta$, $\theta_1$ and $\theta_2$ are the free parameters, $\rho_{AB}$ is a two-qubit state subject to pooling, and $U^{\dagger}_{p}$ is the conjugate transpose of the unitary gate for pooling. The pooling operation in Eq.~(\ref{eq:ZXpooling}) is referred to as ZX pooling. In addition to ZX pooling, generalized pooling is introduced as
\begin{equation}
    \Tr_A\left\lbrack\left(\ketbra{1}{1}_A\otimes U(\theta_1,\phi_2,\lambda_3)_B + \ketbra{0}{0}_A\otimes U(\theta_4,\phi_5,\lambda_6)_B\right) \rho_{AB} U^{\dagger}_{p} \right\rbrack.
\end{equation}
Here, $U(\theta,\phi,\lambda) = R_z(\phi)R_x(-\pi/2)R_z(\theta)R_x(\pi/2)R_z(\lambda)$ and can implement any unitary operator in $SU(2)$. Again, $U^{\dagger}_{p}$ is the conjugate transpose of the corresponding unitary gate for pooling. The unitary gates used in ZX pooling and generalized pooling are compared in Fig.~\ref{fig:pooling}.

\subsection{Transfer learning}
Transferring the knowledge accumulated from one task to another is a typical intelligent behavior that human learners always experience. TL refers to the application of this concept in ML. Specifically, TL aims to improve the training of a new ML model by utilizing a reference (or source) ML model that is pre-trained for a different but related task with a different dataset~\cite{Bozinovski2020ReminderOT,5288526,10.1007/978-3-030-01424-7_27,Goodfellow-et-al-2016}. 
Transfer learning encompasses three main categories: inductive transfer learning (ITL), transductive transfer learning (TTL), and unsupervised transfer learning (UTL)~\cite{5288526, comprehensive_survey_TL_2021}. ITL applies when label information is available for the target domain, while TTL applies when label information is only available for the source domain. UTL, on the other hand, applies when label information is unavailable for both the source and target domains. In this study, we have chosen to focus exclusively on ITL to ensure simplicity and clarity in our explanations and demonstrations. Henceforth, when we refer to transfer learning, it pertains specifically to ITL. Detailed information regarding our numerical simulations will be presented later in the manuscript.

TL is known to be particularly useful for training a deep learning model that takes a long time owing to the large amount of data, especially if the features extracted in early layers are generic across various datasets. In such cases, starting from a pre-trained network such that only a portion of the model parameters is fine-tuned for a particular task can be more practical than training the entire network from scratch. For example, suppose that a neural network is trained with data $A$ to solve task $A$ and finds the set of parameters (i.e., weights and biases) $\boldsymbol{w}_{A}\in\mathbb{R}^{N_A}$. To solve task $B$ given dataset $B$, the neural network is not trained from scratch, as this may require vast computational resources. Instead, the parameters associated with some of the earlier layers of the reference neural network are used as a set of fixed parameters for the new neural network that is subjected to solving task $B$ with data $B$. In other words, some elements of the parameters for this new learning problem, denoted by $\boldsymbol{w}_{B}\in\mathbb{R}^{N_B}$, are identical to those of $\boldsymbol{w}_{A}$. Hence the number of parameters subject to new optimization is less than $N_B$. The successful application of TL can improve training performance by starting from a higher training accuracy, achieving a faster rate of accuracy improvement, and converging to a higher asymptotic training accuracy~\cite{10.5555/1803899}.

The aforementioned observations imply that TL is also beneficial when the amount of data available is insufficient to work with or extremely small to build a good model. Because processing big data in the NISQ era will be challenging, working with small amounts of data through TL is a promising strategy for near-term quantum ML. The target ML model subjected to fresh training (i.e., fine-tuning) in TL typically has a much smaller number of parameters than the pre-trained model. This and the success of QCNN in the few-parameter regime together promote the development of the classical-to-quantum CNN transfer learning.

\section{Classical-to-quantum transfer learning}
\label{sec:c2q}

An extension of TL to quantum ML was proposed, and its general concept was formulated in Ref.~\cite{Mari2020transferlearning}. Although the performances of the quantum models were not compared with those of their classical counterparts, three different scenarios of quantum TL, namely C2Q, quantum-to-classical, and quantum-to-quantum, were shown to be feasible. Among these three possible scenarios, we focus on C2Q TL as mentioned in Section~\ref{sec:intro}, because we aim to utilize QCNNs in the few-parameter regime to the full extent.
% because this is a viable option in the NISQ era in which training a quantum model can be costlier than training a classical model. 
% Moreover, 
Sufficient reduction of the data dimensionality (i.e., the number of attributes or features) by classical learning would ensure that the size of a quantum circuit subject to training is sufficiently small for implementation with NISQ devices. The dimensionality reduction technique is also necessary to simplify expensive quantum state preparation routines to represent classical data in a quantum state~\cite{PhysRevA.64.014303,PhysRevA.73.012307,Mosca:01,PhysRevA.83.032302,Mottonen:2005:TQS:2011670.2011675,araujo_divide-and-conquer_2021,9259210,PhysRevA.102.032420,araujo2021configurable}.

C2Q TL has been utilized for image data classification~\cite{Mari2020transferlearning,mogalapalli_classicalquantum_2022} and spoken command recognition~\cite{qi2021classical}. These works serve as proof of principle for the general idea and present interesting examples to motivate further investigations and benchmarks. The parameterized quantum circuits therein are vulnerable to the barren plateau problem, because they follow the basic structure of a fully connected feedforward neural network with the same number of input and output qubits. Moreover, these studies used classical neural networks to significantly reduce the number of data features to only four or eight. This means that most of the feature extraction is performed classically; hence, the necessity of the quantum part is unclear. These studies encode the reduced data onto a quantum circuit using simple single-qubit rotations, also known as qubit encoding~\cite{grant_hierarchical_2018,PhysRevA.102.032420}, which makes the number of model parameters grow polynomially with the number of data features. In contrast, the number of model parameters in our ML algorithm scales logarithmically with the number of input qubits. Furthermore, all of these works use only one type of ansatz based on repetitive applications of single-qubit rotations and controlled-NOT gates. Finally, the performance of C2Q TL was not compared with that of the C2C version in any of these studies. Because the pre-trained classical neural network performs a significant dimensionality reduction (and hence feature extraction), the absence of a direct comparison with C2C TL raises the question of whether the quantum model achieves any advantage over its classical counterparts.

\begin{figure}[t]
    \centering
    \includegraphics[width=0.96\textwidth]{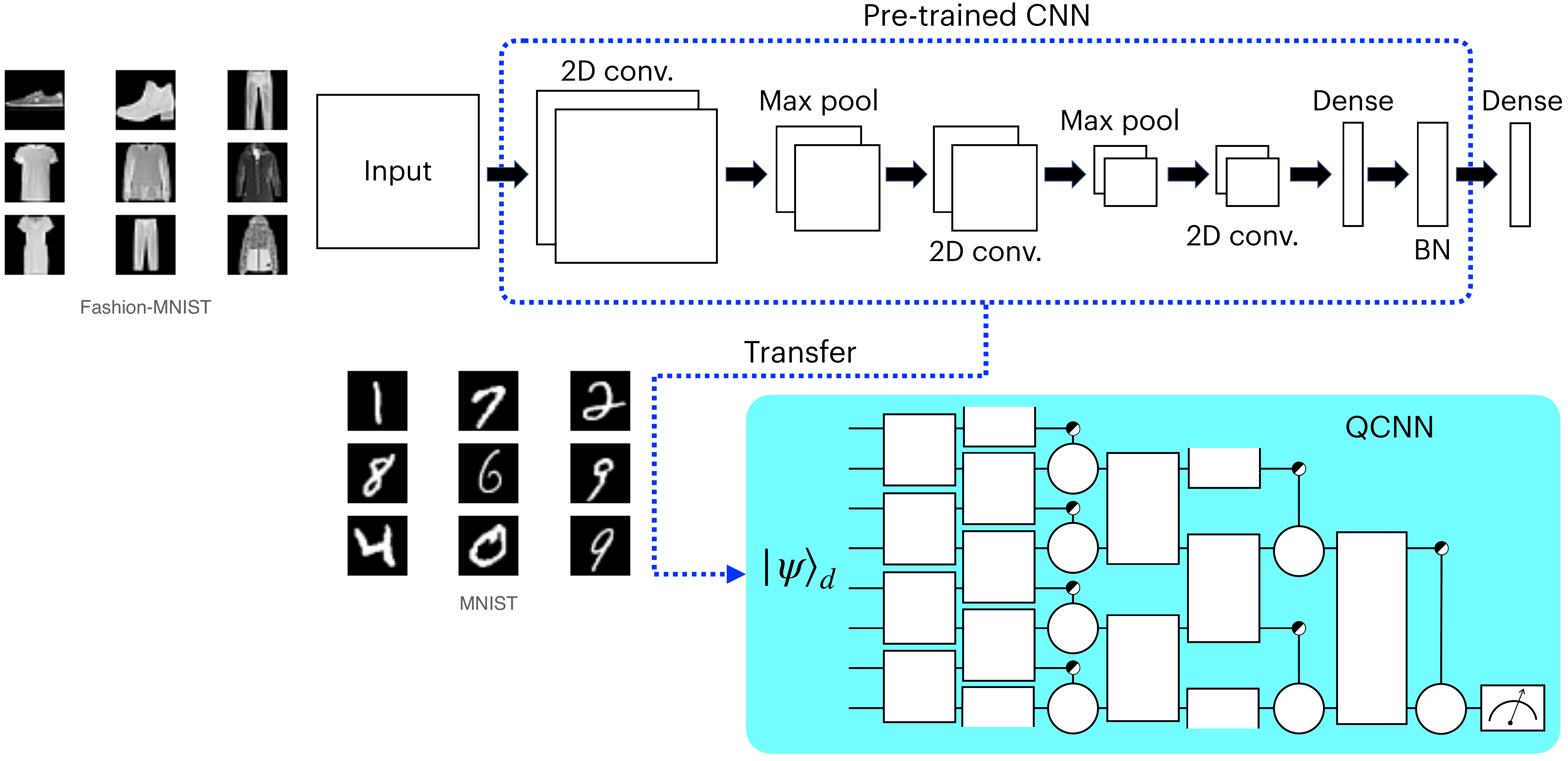}
    \caption{An example of classical-to-quantum convolutional neural network transfer learning simulated in this work for benchmarking and comparison to purely classical models. A source CNN is trained on the Fashion-MNIST dataset. Then, the transfer learning trains a QCNN for MNIST data classification by utilizing the earlier layers of the pre-trained CNN for feature extraction. The source CNN contains convolution (conv.), pooling, dense and batch normalization (BN) layers. 
    }
    \label{fig:4}
\end{figure}

In this study, we present a classical-to-quantum transfer learning framework with QCNN. Our framework facilitates the transfer of knowledge from a pre-trained classical CNN to a quantum CNN, leveraging the unique advantages offered by QCNNs. The adoption of QCNNs as the target model holds crucial importance for several reasons. Firstly, QCNNs possess the capability to circumvent the barren plateau effect, a critical bottleneck encountered during the training of quantum neural networks. This property of QCNNs addresses a major challenge in quantum machine learning and enhances the training process. Furthermore, previous research has demonstrated the advantages of QCNNs over their classical counterparts, particularly in the few-parameter regime, along with their good generalization capabilities. As a result, fine-tuning a machine learning model using a QCNN is expected to yield enhanced classification performance compared to fine-tuning with a traditional CNN. To illustrate the practical implementation of our C2Q TL framework, we provide an example of transfer learning using C2Q-CNN, which serves as the basis for our benchmark studies. The schematic representation of this process can be found in Fig.~\ref{fig:4}, showcasing the application of our proposed framework.

% In this study, we presents a C2Q TL framework with QCNN, which transfers knowledge from a pre-trained classical CNN to a quantum CNN. The use of QCNN as the target model is crucial for several reasons. First, QCNN can avoid the barren plateau effect, which is a critical bottleneck in training a quantum neural network. Moreover, previous studies have shown the advantages of QCNNs over their classical counterpart in the few-parameter regime and its excellent generalization capabilities. Thus, fine-tuning an ML model with a QCNN is expected to improve the classification performance compared to fine-tuning it with a CNN. An example of TL using C2Q-CNN, which was used in our benchmark studies, is illustrated in Fig.~\ref{fig:4}.
% % In this study, we developed C2Q TL with a QCNN and compared C2Q TL results with C2C TL. 
% In this study, we developed C2Q TL with a QCNN (C2Q-CNN), which transfers knowledge from pre-learned (source) model to QCNN.
% Transferring pre-learned information to QCNN is important because QCNN can avoid the barren plateau effect. Moreover, because previous studies have shown the advantages of QCNNs over their classical counterpart in the few-parameter regime, fine-tuning an ML model with a QCNN is expected to improve the classification performance compared to fine-tuning it with a CNN.
% An example of TL using C2Q-CNN, which was used in our benchmark studies, is illustrated in Fig.~\ref{fig:4}.
The general model is flexible with the choice of data encoding, which loads classical data features to a quantum state $|\psi\rangle_d$, and ansatz, the quantum circuit model subject to training. We performed extensive benchmarking over the various ansatzes presented in Section~\ref{sec:qcnn} to classify MNIST data using a classical model pre-trained with Fashion-MNIST data. Finally, we compared the classification accuracies of C2Q and various C2C models. The C2Q models performed noticeably better than all C2C models tested in this study under similar training conditions. More details on the simulation and results are presented in the following section.

\section{Simulation Results}
\label{sec:results}

To demonstrate the advantage of C2Q-CNN, we performed classical simulations of binary classification using PennyLane~\cite{bergholm2020pennylane}. The benchmark was performed using two standard image datasets, MNIST and Fashion-MNIST, which were accessed through Keras~\cite{chollet2015keras}. Examples of the datasets are shown on the left in Fig.~\ref{fig:4}. Note that both datasets have 28$\times$28 features and 10 classes. Among the 10 classes of MNIST data, we performed three independent binary classification tasks aimed at distinguishing between 0 and 1, between 2 and 3, and between 8 and 9. To represent classical data as a quantum state in a QCNN, the classical data must be encoded into a quantum state. The number of data features that can be encoded in $N$ qubits ranges from $N$ to $2^N$ depending on the choice of the encoding method~\cite{grant_hierarchical_2018,PhysRevA.102.032420,Havlicek2019,araujo_divide-and-conquer_2021,araujo2021configurable}. Among many options, we used amplitude encoding to represent as many features as possible (see ~\ref{sec:encoding}). All C2Q-CNN simulations were performed with eight input qubits, to which the amplitude encoding loads 256 features. Quantum feature maps that encode only eight features, such as qubit encoding, are not considered because they require extreme dimensionality reduction on the classical end, which may dominate the classification result as described in the previous section. Furthermore, amplitude encoding was shown to work well with QCNNs for classical data classification~\cite{hur2021quantum}.
% As encoding methods, we provide amplitude encoding and qubit encoding in~\ref{sec:encoding}. We used eight qubits to benchmark the C2Q-CNN. Therefore, amplitude encoding encodes 256 features into eight qubits, whereas qubit encoding encodes only eight features. Only amplitude encoding was used in this simulation due to the small feature space of qubit encoding.

% Traditional CNNs have shown remarkable performance in image classification tasks due to their ability to capture important characteristics of images, such as the stationarity of statistics and the locality of pixel dependencies~\cite{alexnet_2017, resnet, densenet}.
% CNNs have demonstrated remarkable performance in image classification, owing to their almost accurate assumptions about the characteristic of images, such as stationarity of statistics and locality of pixel dependencies~\cite{alexnet_2017, ResNet, DenseNet}. To effectively handle the MNIST and the Fashion-MNIST datasets in our study, we utilized traditional CNNs as the source model, as they have been proven to be highly suitable for these benchmark datasets.
The source classical CNN model, depicted in Fig.~\ref{fig:4}, was trained on 60,000 Fashion-MNIST data for multinomial classification. The TL is implemented by replacing the final dense layer of the pre-trained CNN with a QCNN. The pre-trained CNN is utilized as the source model to perform binary classification on the MNIST dataset. The source model takes the MNIST data as input and outputs 256 values using the parameters pre-trained on the Fashion-MNIST dataset. These output values are then encoded into 8 input qubits of the QCNN, which is fine-tuned using 10,000 MNIST data samples. The fine-tuning process adjusts the parameters of the QCNN to optimize its performance on the binary classification task. The trainable parameters in the QCNN were optimized by minimizing the cross-entropy cost function with the Adam optimizer \cite{kingma2014adam} using PennyLane \cite{bergholm2020pennylane}. The number of MNIST test data was approximately 2,000.

C2Q transfer models can be split into three sets based on different pooling variations. The first set uses ZX pooling with convolution circuits (a)-(j), as shown in Fig.~\ref{fig:convolution}. The second set includes the generalized pooling circuit and convolution circuits (a)-(j), as shown in Fig.~\ref{fig:convolution}. Finally, we constructed transfer models without parameterized quantum gates in pooling layers. We refer to this pooling strategy that merely traces out one of the qubits as trivial pooling.
Trivial pooling is tested with convolution circuits (j) and (k), as shown in Fig.~\ref{fig:convolution}.

To compare C2Q-CNN TL classification results with its classical counterparts, 1D and 2D CNN C2C TL models were constructed with a similar number of trainable parameters as C2Q-CNN models.
The 1D CNN model was composed of a 1D convolution layer and 1D max pooling layer with 64 trainable parameters. Similarly, the 2D CNN model was composed of a 2D convolution layer and 2D max pooling layer with 76 trainable parameters. The CNNs subjected to fine-tuning for the MNIST data use the cross-entropy cost function with the Adam optimizer, as in the C2Q-CNN case. These classical CNN architectures are built using Keras~\cite{chollet2015keras}. Detailed descriptions of C2C models are provided in~\ref{appendix:cc_transfer_learning}.
% During the training process, a small batch was created from the training data at each iteration.
% Compared to training on the full dataset, training on the mini-batch not only reduces the simulation time, but also helps the gradients to escape from local minima~\cite{keskar2016large}.
The training process used mini-batch gradient descent with a batch size of 50 and a learning rate of 0.01. We also fixed the number of training iterations at 200 for the C2Q TL and C2C TL models. The other training conditions were kept the same in the C2Q and C2C transfer models to make the comparison as fair as possible.

\begin{figure}[ht]
    \centering
    \includegraphics[width=1\textwidth]{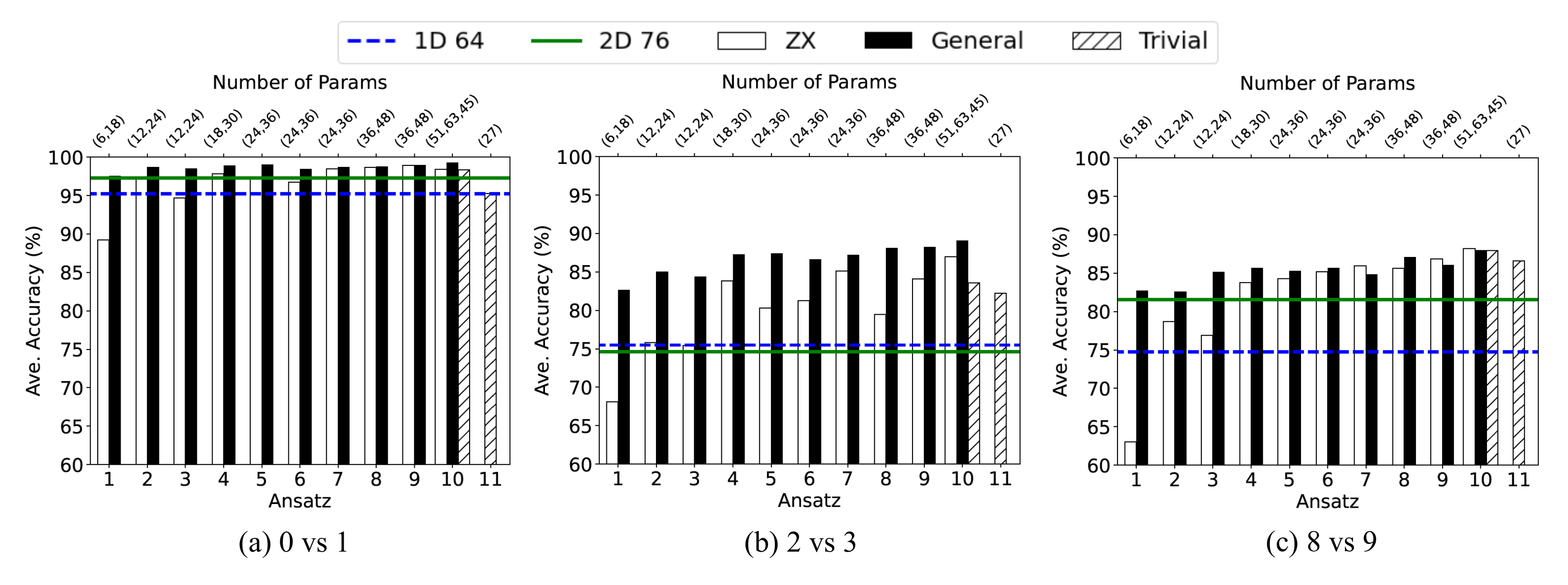}
    \caption{Summary of the classification results with PennyLane simulations (quantum part) and Keras (classical part). Each bar represents the classification test accuracy of C2Q TL averaged over 10 instances given by the random initialization of parameters. The different bars along the x-axis indicate that the results are for different convolution ansatz, labeled according to Fig.~\ref{fig:convolution}. The unfilled, filled, and hatched bars represent the results of ZX pooling, generalized pooling, and trivial pooling, respectively. The number of trainable model parameters for each case is shown at the top of the x-axis. The horizontal lines represent the results of the C2C TL with 1D and 2D CNN architectures. The number of trainable model parameters for each case is provided in the legend.}
    \label{fig:main}
\end{figure}

The TL classification results are shown in Fig.~\ref{fig:main}.
Each bar represents the C2Q classification accuracy averaged over ten randomly initialized parameters.
Different bars along the x-axis represent different convolutions, labeled according to Fig.~\ref{fig:convolution}.
The unfilled, filled, and hatched bars represent the results of ZX pooling, generalized pooling, and trivial pooling, respectively.
The blue dashed line and green solid line represent the results of the C2C TL using 1D and 2D CNN architectures, respectively. 
%A full list of simulation results for each of the ten random initialization instances is provided in~\ref{appendix:result_table}.

The 0 vs. 1 classification results are shown in Fig.~\ref{fig:main} (a). 
For ZX and generalized pooling, most of the average test accuracies were greater than 95\% and 98\%, respectively. The test accuracy with ZX pooling with convolution 2, 4, 7, 8, 9, and 10 ansatz and generalized pooling with all convolution ansatz is greater than that of 1D and 2D C2C TL. The accuracy with trivial pooling with convolution 10 ansatz was also higher than that of C2C TL. In addition, the accuracy of trivial pooling with convolution ansatz 10 is higher than that of trivial pooling with convolution ansatz 11. This can be attributed to the fact that the ansatz 10 has more parameters and is capable of expressing an arbitrary $SU(4)$ including the ansatz 11. The test accuracy of generalized pooling is greater than that of ZX pooling when the convolution ansatz is the same. This can be inferred from the fact that generalized pooling can be trained to learn ZX pooling.

The 2 vs. 3 classification results are shown in Fig.~\ref{fig:main} (b). Most of the ZX pooling average accuracies were between 70\% and 90\%, and most of the generalized pooling average accuracies were between 85\% and 90\%. The test accuracies in (b) are lower than those of (a) because 2 vs. 3 image classifications are more difficult than 0 vs. 1 image classifications. All C2Q TL classification accuracies are higher than the C2C classification accuracies except for ZX pooling with convolution 1 and 3, which use a much smaller number of parameters than the purely classical TL. The test accuracy of generalized pooling is greater than that of ZX pooling when the convolution is the same. In addition, the accuracy of trivial pooling with convolution ansatz 10 is higher than that with ansatz 11. These results are consistent with those obtained from the 0 vs. 1 classification, providing further evidence that the ansatz with more model parameters and improved expressibility is favored. This suggests that the use of more complex models, such as the generalized pooling, is advantageous in solving classification problems, particularly in comparison to less expressive models such as ZX pooling.

The 8 vs. 9 classification results are shown in Fig.~\ref{fig:main} (c). Most of the ZX pooling average accuracies were between 70\% and 90\%, and most of the generalized pooling average accuracies were between 85\% and 90\%. The test accuracies in (c) are lower than those in (a) because 8 vs. 9 image classification is more difficult than 0 vs. 1 image classification, but the accuracies in (c) are similar to those in (b). All C2Q TL classification accuracies are higher than the C2C TL classification accuracies except for ZX pooling with convolution 1, 2, and 3, which use a much smaller number of parameters than the purely classical TL. As before, trivial pooling with convolution ansatz 10 has higher accuracy than trivial pooling with ansatz 11. 
the accuracy of generalized pooling is consistently higher than that of ZX pooling when the underlying convolution is identical. 
However, there are instances where the performance of ZX pooling is comparable to that of generalized pooling, such as with the use of convolution ansatz 7. Nevertheless, a Welch's t-test analysis~\cite{welch_t_test} revealed that these results are not statistically significant. Based on these findings, we can conclude that generalized pooling is a more favorable approach for solving all of the classification problems tested, compared to ZX pooling.

The results in Fig.~\ref{fig:main} (a), (b), and (c) show a QCNN's tendency to perform better when the convolution circuits have a larger number of trainable parameters. However, simply increasing the number of trainable parameters does not always guarantee to improve the test accuracy because it is affected by various conditions, such as statistical error and quantum gate arrangement. For example, ZX pooling with convolution ansatz 5, 6, 7 in (b) has the same number of trainable parameters, but their average accuracies are different. In the current study, overfitting was not observed as the number of model parameters was much smaller than the number of data, but it is important to keep this issue in mind when designing and implementing larger models in the future.

In summary, generalized pooling mostly produces higher classification accuracy than ZX pooling with the same convolution circuit. This is as expected since the generalized pooling has more model parameters. Moreover, it can be reduced to ZX pooling under appropriate parameter selection. The accuracy of all ZX pooling, generalized pooling, and trivial pooling circuits tends to be higher when the convolution circuits have a larger number of gate parameters. Although C2Q models have fewer trainable parameters than C2C models, most C2Q models outperform C2C models.

To validate our findings, we conducted a Welch's t-test analysis~\cite{welch_t_test} to determine the statistical significance of the improved classification results obtained by the C2Q models. Our results show that ZX pooling with convolution ansatz 9 and generalized pooling with convolution ansatz 4, 5, 9, and 10 have a statistically significant quantum advantage over both 1D and 2D classical models for all classification problems, despite having a smaller number of model parameters. Further details on the statistical analysis can be found in \ref{appendix:p-value}. These findings underscore the potential of quantum-enhanced machine learning models in solving complex classification tasks, even with limited model resources.

The underlying source of the quantum advantage in quantum computing remains an open question. However, it is speculated that the advantage is related to certain properties of quantum computing that have no classical equivalent. The first property is the ability of quantum measurements to discriminate non-orthogonal states, which enables quantum computers to capture subtle differences in data that are not captured by classical computers. The second property is the ability of quantum convolutional operations to create entanglement among all qubits through the use of two-qubit gates between nearest neighbors, which allows for the capture of non-local correlations. In addition, the ability of a quantum computer to store $N$-dimensional data in $\lceil\log_2(N)\rceil$ qubits, and the ability of the QCNN to classify $M$-qubit quantum states using only $O(\log(M))$ parameters, make it possible to construct an extremely compact machine learning model.

\section{Conclusion}
\label{sec:conclusion}
In this study, we proposed a classical-to-quantum CNN (C2Q-CNN), a transfer learning (TL) model that uses some layers of a pre-trained CNN as a starting point for a quantum CNN (QCNN). The QCNN constitutes an extremely compact machine learning (ML) model because the number of trainable parameters grows logarithmically with the number of initial qubits~\cite{cong_quantum_2019} and is promising because of the absence of barren plateaus~\cite{pesah2020absence} and generalization capabilities~\cite{PRXQuantum.2.040321}. Supervised learning with a QCNN has also demonstrated classification performance superior to that of its classical counterparts under similar training conditions for a number of canonical datasets~\cite{hur2021quantum}. C2Q-CNN TL provides an approach to utilize the advantages of QCNN in the few-parameter regime to the full extent. Moreover, the proposed method is suitable for implementation in quantum hardware expected to be developed in the near future because it is robust to systematic errors and can be implemented with a shallow-depth quantum circuit. Therefore, C2Q-CNN TL is a strong candidate for practical applications of NISQ computing in ML with a quantum advantage.

To demonstrate the quantum advantage of C2Q-CNN, we conducted a comparative study between two classical-to-classical (C2C) transfer learning (TL) models and C2Q TL models. The C2C and C2Q TL models shared the same pre-trained CNN, with the C2C TL models having slightly more parameters than the C2Q TL models. The pre-training was performed on the Fashion-MNIST dataset for multinomial classification. Then the target model replaced the final dense layer of the source model and was trained for three independent binary classification tasks using the MNIST data. Our simulation results, obtained using PennyLane and Keras, revealed that the C2Q models consistently achieved higher classification accuracy compared to the C2C models, despite having fewer trainable parameters. These results highlight the potential of quantum-enhanced transfer learning in improving the performance of machine learning models. It is important to note that while our simulation utilized a source CNN specifically designed for this study, the C2Q-CNN TL framework is compatible with other existing CNNs, such as VGGNet~\cite{VGGNet}, ResNet~\cite{ResNet}, and DenseNet~\cite{DenseNet}. This compatibility enhances the versatility of our approach, enabling researchers to leverage established CNN designs within the quantum-enhanced transfer learning paradigm.

The potential future research directions are as follows. First, the reason behind the quantum advantage demonstrated by C2Q-CNN remains unclear. Although rigorous analysis is lacking, we speculate that this advantage is related to the ability of a quantum measurement to discriminate non-orthogonal states, for which a classical analog does not exist. Moreover, verifying whether the quantum advantage would continue to hold as the number of trainable parameters increases and for other datasets would be interesting. To increase the number of model parameters for a fixed number of features and input qubits, one may consider generalizing the QCNN model to utilize multiple channels, as in many classical CNN models. Note that, in the TL tested in our experiment, the final dense layer was replaced with a model subjected to fine-tuning, while the entire convolutional part was frozen. Testing the various depths of frozen layers would be an interesting topic for future research. For example, freezing a smaller number of layers to use the features of an earlier layer of the convolutional stage can be beneficial when the new dataset is small and significantly different from the source. The focus of this study was on inductive TL, for which both the source and new datasets were labeled. Exploring the potential of leveraging quantum techniques in other TL scenarios, such as self-taught, unsupervised, and transductive TL~\cite{5288526}, is a promising direction for future research. Furthermore, extending the C2Q TL approach to address other machine learning problems, such as semi-supervised learning~\cite{semi_book, semi_paper_2022} and one-class classification~\cite{deep_occ, occ_paper_2022}, poses an open challenge for future research.

\section*{Data Availability}
The data that support the findings of this study, including additional benchmarking examples not explicitly mentioned in the manuscript, are available upon request.

\section*{Acknowledgments}

This research was supported by the Yonsei University Research Fund of 2022 (2022-22-0124), the National Research Foundation of Korea (Grant Nos. 2021M3H3A1038085, 2019M3E4A1079666, 2022M3E4A1074591, 2022M3H3A1063074 and 2021M3E4A1038308), and the KIST Institutional Program (2E32241-23-010). 

\appendix
\section{Encoding classical data to a quantum state}
\label{sec:encoding}

The first step in applying quantum ML to a classical dataset is to transform the classical data into a quantum state. Without loss of generality, we consider the classical data given as an $N$-dimensional real vector $\vec{x}\in\mathbb{R}^N$. Several encoding methods exist to achieve this, such as algorithms that require a quantum circuit with $O(N)$ width and $O(1)$ depth and algorithms that require $O(\log(N))$ width and $O(poly(N))$ depth~\cite{araujo_divide-and-conquer_2021,9259210,PhysRevA.102.032420,araujo2021configurable}.
Among the various encoding methods explored previously~\cite{hur2021quantum}, we observed that amplitude encoding performs best in most cases. 
\subsection{Amplitude encoding}
Amplitude encoding encodes classical data into the probability amplitude of each computational quantum state. Amplitude encoding transforms $x=(x_1,...,x_N)^{\top}$ of dimension $N=2^n$ classical data into an n-qubit quantum state $\ket{\psi (x)}$ as follows:
\begin{equation}
    U(x): x \in \mathbb{R}^N \xrightarrow[]{} \ket{\psi (x)} = \frac{1}{||x||} \sum_{i=1}^{N}x_i\ket{i},
\end{equation}
where $\ket{i}$ denotes the $i$th computational basis state. Amplitude encoding can optimize the number of parameters on the $O(log(N))$ scale. However, the quantum circuit depth of amplitude encoding typically increases with $O(poly(N))$.

\subsection{Qubit encoding}
Qubit encoding uses a constant quantum circuit depth, while using $N$ qubits. Qubit encoding rescales classical data to $x_i$ which lies between 0 and $\pi$, and then inputs $x_i$ into a single qubit as $\ket{\psi (x)} = \cos{(\frac{x_i}{2})}\ket{0} + \sin{(\frac{x_i}{2})}\ket{1}$ for $i = 1,...,N$.
Therefore, qubit encoding transforms $x=(x_1,...,x_N)^{\top}$ into $N$ qubits as 
\begin{equation}
    U(x) : x \in \mathbb{R}^N \xrightarrow{} \ket{\psi (x)} = \bigotimes_{i=1}^{N}\bigg(\cos{\left(\frac{x_i}{2}\right)}\ket{0} + \sin{\left(\frac{x_i}{2}\right)}\ket{1}\bigg)
\end{equation}
where $x_i \in [0,\pi)$ for all $i$. This unitary operator $U(x)$ can be expressed as the tensor product of a single qubit unitary operator $U(x) = \otimes_{j=1}^{N} U_{x_j}$, where
\begin{equation}
    U_{x_j} = e^{-i\frac{x_j}{2}\sigma_y} = 
    \begin{bmatrix}
        \cos{\left(\frac{x_j}{2}\right)} & -\sin{\left(\frac{x_j}{2}\right)}\vspace{1mm} \\ 
        \sin{\left(\frac{x_j}{2}\right)} & \cos{\left(\frac{x_j}{2}\right)}
   \end{bmatrix}.
\end{equation}

\section{Classical Neural Network}
\label{appendix:cc_transfer_learning}
We devised classical convolutional neural networks to compare C2C TL and C2Q TL under the same training conditions. In particular, we assigned the similar number of model parameters to C2C TL and C2Q TL.
We created 1D and 2D CNN models for the C2C TL using Keras \cite{chollet2015keras}. We used the same pre-trained CNN model with C2Q TL, which is introduced in Fig.~\ref{fig:4}. To compare the test accuracies of C2C and C2Q transfer models under the same conditions, the learning rate, batch size, and number of iterations were fixed, and the Adam optimizer was used.

\subsection{1D CNN}
\begin{figure}[t]
    \centering
    \includegraphics[scale=0.5]{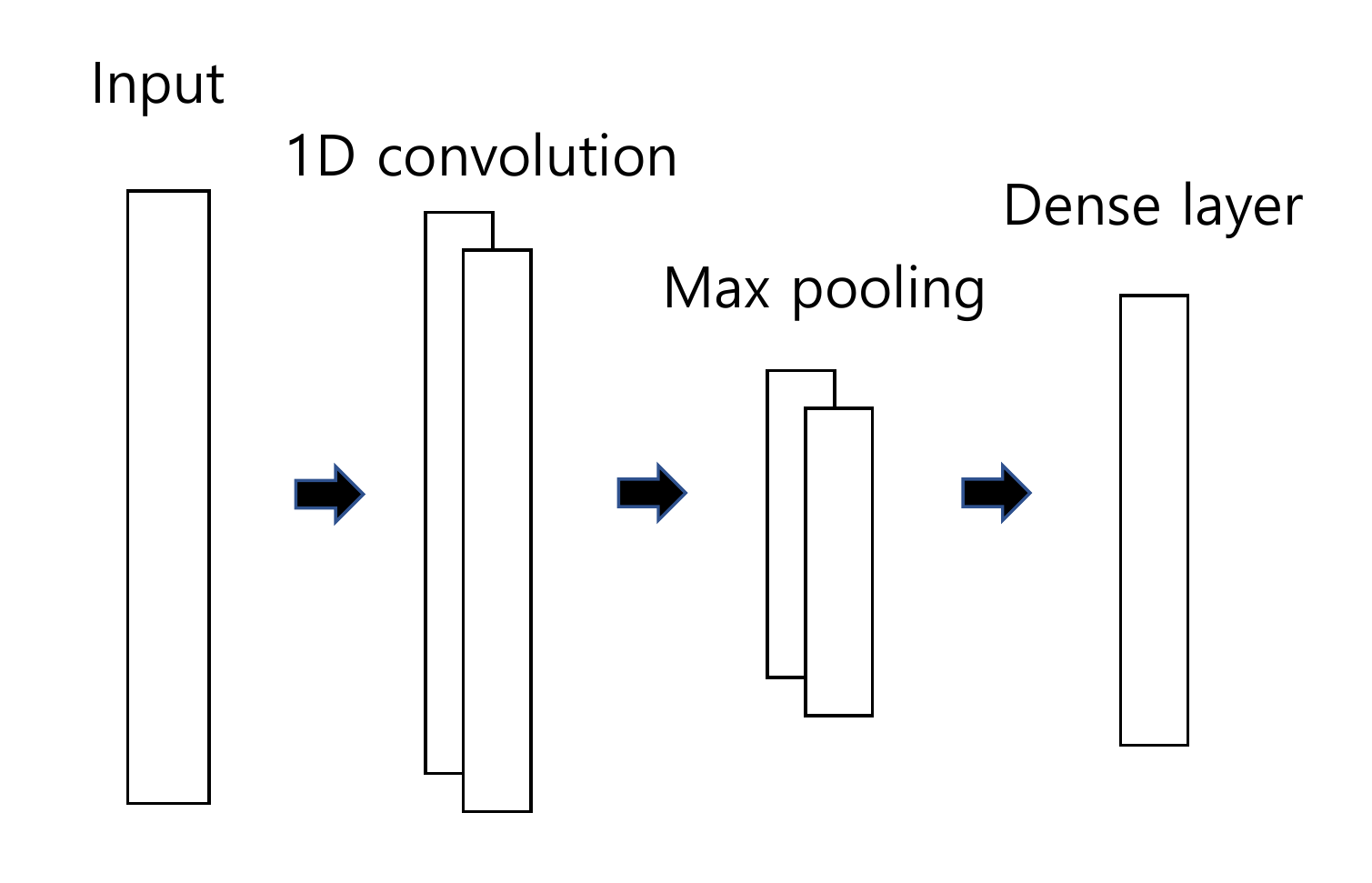}
    \caption{1D CNN model}
    \label{fig:1dcnn}
\end{figure}

The structure of the 1D CNN model is illustrated in Fig.~\ref{fig:1dcnn}. The CNN takes 256 features produced by the source (pre-trained) CNN as input, and passes them on to 1D convolution and max pooling layers. The output feature size is reduced to 28 at the end of the max pooling layer. Finally, a dense layer is applied to reduce the size of the output features to two for binary classification. The total number of trainable parameters is 64. 
%The accuracy test results for these models are presented in \ref{appendix:result_table}.

\subsection{2D CNN}
\begin{figure}[t]
    \centering
    \includegraphics[scale=0.5]{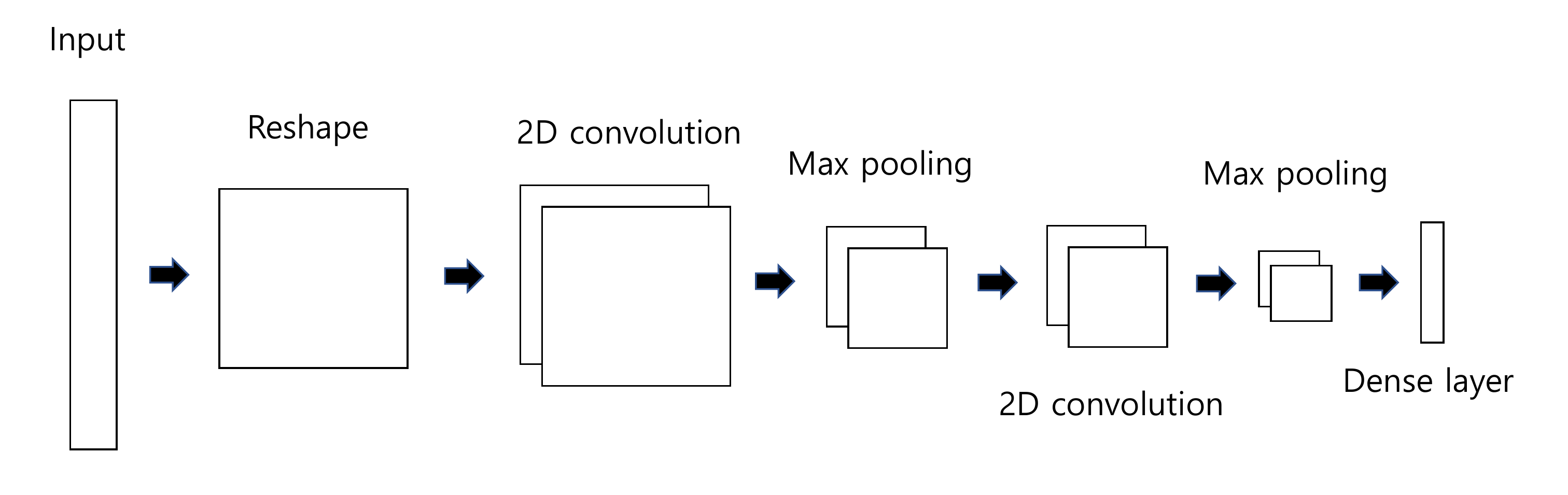}
    \caption{2D CNN model}
    \label{fig:2dcnn}
\end{figure}

The structure of the 2D CNN model is shown in Fig.~\ref{fig:2dcnn}. The CNN takes $16 \times 16$ data, which is reshaped from 256 features produced by the source (pre-trained) CNN as input. This two-dimensional data is passed on to the 2D convolution and max pooling layer twice, and the output features are reduced to eight. Finally, a dense layer is applied to reduce the size of the output features to two for binary classification. The total number of trainable parameters is 76. 
%The accuracy test results for these models are presented in \ref{appendix:result_table}.

\section{Welch's t-test}
\label{appendix:p-value}
The Welch's t-test~\cite{welch_t_test} is a widely-utilized method for assessing the equality of means between two populations with unequal variances. In our study, the Welch's t-test was implemented using SciPy~\cite{2020SciPy-NMeth} to obtain $p$-values between the C2Q TL model and the C2C TL models. In accordance with standard statistical practice~\cite{johnson_statistical_evidence_2013,significance_2013}, a $p$-value less than $\alpha$, where $\alpha$ is typically set to 0.05, is considered to indicate statistical significance. Based on the obtained $p$-values, we conclude that if a C2Q TL model demonstrates statistical significance compared to the C2C TL models and achieves a higher accuracy, it can be inferred to possess a meaningful advantage (quantum advantage). The $p$-value results were organized into groups based on the ZX pooling, generalized pooling, and trivial pooling approaches, as previously discussed in the paper. For each pooling type, the $p$-value results were further grouped by the type of learning problem, which in this case was binary classification for 0 and 1, 2 and 3, and 8 and 9. The results of ZX pooling are presented in Tables \ref{table:zx_01_p}, \ref{table:zx_23_p}, and \ref{table:zx_89_p} for classifying between 0 and 1, 2 and 3, and 8 and 9, respectively. The results of generalized pooling are presented in Tables \ref{table:g_01_p}, \ref{table:g_23_p}, and \ref{table:g_89_p} in the same order. Finally, the results of trivial pooling are listed in Table~\ref{table:trivial_p}. We highlighted in bold any statistically significant $p$-values where the corresponding C2Q model exhibited higher accuracy than the C2C model. 
% All statistically significant p-values are highlighted in bold in the tables. 

\newcommand{\bfsel}{\fontseries{b}\selectfont}

\begin{table}[h!]
    \centering
    \begin{tabular}{|l|l|l|l|l|l|l|l|l|l|l|}
    \hline
        convolution circuit & 1 & 2 & 3 & 4 & 5 & 6 & 7 & 8 & 9 & 10  \\ \hline
        1D 64 $p$-value & 0.0312  & 0.1289  & 0.7327  & 0.0793  & 0.1795  & 0.2957  & \bfsel 0.0316  & \bfsel 0.0240  & \bfsel 0.0166  & \bfsel 0.0422   \\ \hline
        2D 76 $p$-value & 0.0047  & 0.8761  & 0.0323  & 0.5240  & 0.9725  & 0.5402  & 0.1348  & 0.0888  & \bfsel 0.0353  & 0.2441   \\ \hline
       % Quantum advantage  & ~ & ~ & ~ & ~ & ~ & ~ & ~ & ~ & TRUE  & ~ \\ \hline
    \end{tabular}
    \caption{ZX pooling with the 0 vs. 1 classification $p$-value results.}
    \label{table:zx_01_p}
\end{table}

\begin{table}[h!]
    \centering
    \begin{tabular}{|l|l|l|l|l|l|l|l|l|l|l|}
    \hline
        convolution circuit & 1 & 2 & 3 & 4 & 5 & 6 & 7 & 8 & 9 & 10  \\ \hline
        1D 64 $p$-value & 0.0768  & 0.9352  & 0.9927  & \bfsel 0.0497  & 0.2332  & 0.1700  & \bfsel 0.0304  & 0.3209  & \bfsel 0.0446  & \bfsel 0.0119   \\ \hline
        2D 76 $p$-value & 0.0289  & 0.6896  & 0.7954  & \bfsel 0.0047  & 0.0579  & \bfsel 0.0375  & \bfsel 0.0023  & 0.0985  & \bfsel 0.0040  & \bfsel 0.0006   \\ \hline
       % Quantum advantage  & ~ & ~ & ~ & TRUE  & ~ & ~ & TRUE  & ~ & TRUE  & TRUE  \\ \hline
    \end{tabular}
    \caption{ZX pooling with the 2 vs. 3 classification $p$-value results.}
    \label{table:zx_23_p}
\end{table}

\begin{table}[h!]
    \centering
    \begin{tabular}{|l|l|l|l|l|l|l|l|l|l|l|}
    \hline
        convolution circuit & 1 & 2 & 3 & 4 & 5 & 6 & 7 & 8 & 9 & 10  \\ \hline
        1D 64 $p$-value & 0.0008  & 0.2373  & 0.4708  & \bfsel 0.0035  & \bfsel 0.0024  & \bfsel 0.0012  & \bfsel 0.0007  & \bfsel 0.0010  & \bfsel 0.0004  & \bfsel 0.0002   \\ \hline
        2D 76 $p$-value & 0.0000  & 0.2806  & 0.0495  & 0.1591  & 0.0904  & \bfsel 0.0204  & \bfsel 0.0117  & \bfsel 0.0085  & \bfsel 0.0015  & \bfsel 0.0002   \\ \hline
       % Quantum advantage  & ~ & ~ & ~ & ~ & ~ & TRUE  & TRUE  & TRUE  & TRUE  & TRUE  \\ \hline
    \end{tabular}
    \caption{ZX pooling with the 8 vs. 9 classification $p$-value results.}
    \label{table:zx_89_p}
\end{table}

\begin{table}[h!]
    \centering
    \begin{tabular}{|l|l|l|l|l|l|l|l|l|l|l|}
    \hline
        convolution circuit & 1 & 2 & 3 & 4 & 5 & 6 & 7 & 8 & 9 & 10  \\ \hline
        1D 64 $p$-value & 0.1079  & \bfsel 0.0215  & \bfsel 0.0302  & \bfsel 0.0169  & \bfsel 0.0139  & \bfsel 0.0297  & \bfsel 0.0217  & \bfsel 0.0206  & \bfsel 0.0151  & \bfsel 0.0103   \\ \hline
        2D 76 $p$-value & 0.7661  & 0.0535  & 0.1171  & \bfsel 0.0308  & \bfsel 0.0197  & 0.1026  & 0.0598  & 0.0609  & \bfsel 0.0268  & \bfsel 0.0107   \\ \hline
        % Quantum advantage  & ~ & ~ & ~ & TRUE  & TRUE  & ~ & ~ & ~ & TRUE  & TRUE  \\ \hline
    \end{tabular}
    \caption{Generalized pooling with the 0 vs. 1 classification $p$-value results.}
    \label{table:g_01_p}
\end{table}

\begin{table}[h!]
    \centering
    \begin{tabular}{|l|l|l|l|l|l|l|l|l|l|l|}
    \hline
        convolution circuit & 1 & 2 & 3 & 4 & 5 & 6 & 7 & 8 & 9 & 10  \\ \hline
        1D 64 $p$-value & 0.0921  & \bfsel 0.0290  & \bfsel 0.0389  & \bfsel 0.0106  & \bfsel 0.0099  & \bfsel 0.0138  & \bfsel 0.0109  & \bfsel 0.0074  & \bfsel 0.0068  & \bfsel 0.0048   \\ \hline
        2D 76 $p$-value & \bfsel 0.0129  & \bfsel 0.0021  & \bfsel 0.0034  & \bfsel 0.0006  & \bfsel 0.0006  & \bfsel 0.0008  & \bfsel 0.0006  & \bfsel 0.0004  & \bfsel 0.0003  & \bfsel 0.0002   \\ \hline
        % Quantum advantage  & ~ & TRUE  & TRUE  & TRUE  & TRUE  & TRUE  & TRUE  & TRUE  & TRUE  & TRUE  \\ \hline
    \end{tabular}\caption{Generalized pooling with the 2 vs. 3 classification $p$-value results.}
    \label{table:g_23_p}
\end{table}

\begin{table}[h!]
    \centering
    \begin{tabular}{|l|l|l|l|l|l|l|l|l|l|l|}
    \hline
        convolution circuit & 1 & 2 & 3 & 4 & 5 & 6 & 7 & 8 & 9 & 10  \\ \hline
        1D 64 $p$-value & \bfsel 0.0080  & \bfsel 0.0079  & \bfsel 0.0013  & \bfsel 0.0009  & \bfsel 0.0012  & \bfsel 0.0010  & \bfsel 0.0016  & \bfsel 0.0004  & \bfsel 0.0007  & \bfsel 0.0002   \\ \hline
        2D 76 $p$-value & 0.4616  & 0.4574  & \bfsel 0.0202  & \bfsel 0.0085  & \bfsel 0.0261  & \bfsel 0.0062  & \bfsel 0.0279  & \bfsel 0.0010  & \bfsel 0.0038  & \bfsel 0.0002   \\ \hline
        % Quantum advantage & ~ & ~ & TRUE & TRUE & TRUE & TRUE & TRUE & TRUE & TRUE & TRUE  \\ \hline
    \end{tabular}
    \caption{Generalized pooling with the 8 vs. 9 classification $p$-value results.}
    \label{table:g_89_p}
\end{table}

\begin{table}[h!]
    \centering
    \begin{tabular}{|l|l|l|l|l|l|l|}
    \hline
        Classification & \multicolumn{2}{|c|}{0 vs 1} & \multicolumn{2}{|c|}{2 vs 3} & \multicolumn{2}{|c|}{8 vs 9}  \\ \hline
        Convolution & 10 & 11 & 10 & 11 & 10 & 11  \\ \hline
        1D 64 $p$-value & \bfsel 0.0340  & 0.9678  & 0.0602  & 0.1097  & \bfsel 0.0002  & \bfsel 0.0005   \\ \hline
        2D 76 $p$-value & 0.1305  & 0.1192  & \bfsel 0.0065  & \bfsel 0.0173  & \bfsel 0.0003  & \bfsel 0.0048   \\ \hline
        % Quantum advantage  & ~ & ~ & ~ & ~ & TRUE  & TRUE  \\ \hline
    \end{tabular}
    \caption{Trivial pooling $p$-value results.}
    \label{table:trivial_p}
\end{table}

\vfill
% %Bibliography
% \bibliographystyle{unsrt}  
% \bibliography{references} 

\end{document}